\documentclass[aps,prl,amsmath, amssymb,nofootinbib, reprint, twocolumn
]{revtex4-2}

\usepackage{graphicx}
\usepackage{dcolumn}
\usepackage{bm}
\usepackage{subcaption}

\usepackage{color}

\begin{document}

\preprint{}

\title{A New Paradigm for Testing Gravity: Theory-Independent Constraints Using Data From All Astrophysical and Cosmological Scales}

\author{Daniel B Thomas}
\email{dan.b.thomas1@gmail.com}
\affiliation{Jodrell Bank Centre for Astrophysics, University of Manchester, UK
}

\author{Theodore Anton}
\email{theoanton123@gmail.com}
\author{Timothy Clifton}
\email{t.clifton@qmul.ac.uk}
\affiliation{
Department of Physics and Astronomy, Queen Mary University of London, UK
}

\date{\today}

\begin{abstract}
Testing General Relativity (GR) is a key science goal of much of modern physics, and usually results in constraints that are either theory or context specific. We present an holistic framework that we dub `Parametrized Post-Newtonian Cosmology' (PPNC), which can be used for obtaining theory-agnostic constraints on deviations from GR using a single unified set of parameters that apply on all astrophysical and cosmological scales. Our approach is based on the formalism and philosophy of the highly successful Parametrized Post-Newtonian (PPN) framework, but allows us to combine observations of the Cosmic Microwave Background (CMB) and Baryon Acoustic Oscillations (BAOs) with Solar System observations of the Cassini probe and ephemeris of Mars. The full combination of these data sets constrains average deviations from GR over the history of the Universe to be less than $\sim 10\%$, with PPNC parameter values $\bar{\alpha}=0.97^{+0.06}_{-0.07}$ and $\bar{\gamma}=0.97^{+0.05}_{-0.05}$ at the $68\%$ confidence level (GR corresponds to $\bar{\alpha}=\bar{\gamma}=1$). We find that these gravitational parameters have a strong mutual degeneracy, and are constrained to be within $1\%$ of each other for all of cosmic history. Our results demonstrate the ability of the PPNC framework to combine astrophysical, Solar System and cosmological tests of gravity into a single set of unified constraints. We expect our approach to be particularly useful for upcoming missions in both cosmology and astrophysics, which ultimately seek to constrain the same underlying gravitational interaction.

 
\end{abstract}

\maketitle


\textit{Introduction}: 
Precision tests of gravity have a long history, starting from Eddington's observation of lensing during a solar eclipse \cite{edd}, and now spanning many observations in both the Solar System and other astrophysical environments \cite{lrr_will} as well as cosmology \cite{Clifton_2012, Ishak_2018}. However, due to the complicated nature of General Relativity (GR) and its alternatives, tests are typically context-specific, and often rely on the specification of theories (or classes of theories). As such, results across disciplines are hard to compare and combine, which precludes a full understanding of (and optimization of constraints on) modifications to Einstein's theory. This issue is particularly acute given the wealth of ongoing and upcoming cosmological and astrophysical observations that can be used to test gravity in some form, such as LSST\footnote{https://www.lsst.org}, Euclid\footnote{https://www.euclid-ec.org}, SO\footnote{https://simonsobservatory.org/}, SKA\footnote{https://www.skatelescope.org}, LISA\footnote{https://lisa.nasa.gov/}, and DESI\footnote{https://www.desi.lbl.gov/}.

Here we use Parametrized Post-Newtonian Cosmology (PPNC) \cite{Sanghai_2017, Sanghai_2019, Anton_2021, Thomas_2023, Thomas_2024, slip}, which is an extension of the Parametrized Post-Newtonian (PPN) approach \cite{Will_1993} to include cosmological background expansion and the evolution  of inhomogeneities. This extends the industry standard formalism for precision tests of gravity in the Solar System into the cosmological domain using a single (extended) set of parameters. Whilst analogous model-agnostic approaches exist for cosmology (see e.g. \cite{Clifton_2012, Ishak_2018} for reviews), they are typically based in cosmological perturbation theory\footnote{See \cite{ps1pf1,ps1pf_forecast} for an exception that uses post-Newtonian methods.}, and therefore cannot be directly related to the PPN parameters. The PPNC approach, on the other hand, has the PPN parameters appearing explicitly in the equations governing the evolution of the universe and structures within it. It is this that allows us to combine cosmological, astrophysical and Solar System data into a single unified set of constraints on deviations from GR, and to constrain the time evolution of the PPN parameters from the early universe through to the present day. We illustrate this using a combination of Cosmic Microwave Background (CMB) \cite{Planck_2020}, Baryon Acoustic Oscillation (BAO) \cite{1106.3366,1409.3238,1409.3242,1607.03155,2007.08991,2007.08993,2007.08994,2007.09009,2007.09008,2007.08998,2007.08999,2007.08995}, Cassini probe \cite{cassini2003} and Mars ephemeris \cite{konopliv2011mars} data. The constraints we obtain represent a natural and robust realization of the idea of a cosmologically-varying Newton's constant, $G$.


\vspace{0.25cm}
\textit{Theory}:
The PPNC approach constructs the cosmological metric as a patchwork of $\sim 100\, {\rm Mpc}$ regions of space, each of which is individually described by the PPN test metric \cite{Will_1993}. It results in
$$
\mathrm{d}s^2=a(\tau)^2\left[-(1-2{\Phi})\mathrm{d}{\tau}^2+(1+2{\Psi})\delta_{ij}\mathrm{d}{x}^i\mathrm{d}{x}^j \right] \, ,
$$
where $a$ is the scale factor, $\tau$ is conformal time and $\Phi$ and $\Psi$ are the cosmological scalar potentials in longitudinal gauge. In a spatially-flat universe\footnote{The formalism permits non-zero spatial curvature \cite{Sanghai_2017}, but we set it to zero in this work.}, dominated by matter and dark energy, the dynamical equations governing the background and inhomogeneities can be written
\begin{eqnarray}
&&\mathcal{H}^2 = \frac{8 \pi G a^2}{3}\, \gamma \, \bar{\rho}-\frac{2a^2}{3} \, \gamma_c
\end{eqnarray}
and
\vspace{-0.7cm}
\begin{eqnarray} \label{pert}
&&\Psi_{,\tau} +\mathcal{H}\Phi=4\pi G a^2  \, \mu \, \bar{\rho} \, v +\mathcal{G} \, \mathcal{H} \Psi \, ,
\end{eqnarray}
with $\Phi=\eta\Psi$. Here, $\mathcal{H} \equiv \dot{a}/a$ is the conformal Hubble rate, $\bar{\rho}$ the mass density of baryons and dark matter, $k$ is the comoving wavenumber, $\delta \equiv \delta \rho/\bar{\rho}$ is the density contrast, and $v$ the velocity potential (such that $v_i\equiv v_{,i}$). The PPN parameters $\alpha=\alpha(a)$ and $\gamma=\gamma(a)$ are now treated as functions of time\footnote{For simplicity we will continue referring to these functions of time as `parameters'.}, with $\alpha_c=\alpha_c(a)$ and $\gamma_c=\gamma_c(a)$ introduced in order to account for dark energy \cite{Sanghai_2017}. The gravitational coupling functions $\mu$, $\mathcal{G}$ and $\eta$ can be written in terms of the PPN parameters on large (L) and small (S) scales as\footnote{`Large' means scales greater than the cosmological horizon, and `small' means below $\sim 100\, {\rm Mpc}$.}
\vspace{-0.25cm}
\begin{eqnarray*}
&\mu_{\rm S}= \gamma \, , \qquad &\mu_{\rm L}= \gamma -\dfrac{1}{3} \hat{\gamma}+\dfrac{1}{12\pi G \bar{\rho}} \, \hat{\gamma}_c\\
&\mathcal{G}_{\rm L}= 0 \, , \qquad &{\mathcal{G}}_{\rm S}= \dfrac{\alpha-\gamma}{\gamma}+ \dfrac{\hat{\gamma}}{\gamma}\,,
\end{eqnarray*}
$\eta_S=\alpha/\gamma$\,, and\footnote{More complicated expressions for $\eta_L$ have little degeneracy with, or impact on, the other parameters \cite{slip}.} $\eta_L=1$. Hats here denote derivatives with respect to the number of e-foldings, i.e. $\hat{X} \equiv \mathrm{d}X/\mathrm{d}\ln \, a$, and we have chosen to set $\Omega_k=0$.

The functional form of the PPN parameters $\alpha(a)$ and $\gamma(a)$ is not specified {\it a priori}, and so we take it to be given by a simple power law:
\vspace{-0.25cm}
$$
X(a)=A\left(\frac{a_1}{a}\right)^n+B \, ,
\vspace{-0.25cm}
$$
which we assume to be valid from our initial time ($a_1=10^{-10}$) until today ($a=1$). Following \cite{Thomas_2023} we take the interpolation of the coupling functions between large and small scales to be given by 
\vspace{-0.25cm}
$$
f(k)=\frac{1}{2} \left( f_{\rm S}+f_{\rm L} \right) + \frac{1}{2} \left( f_{\rm S}-f_{\rm L} \right) \tanh\left( \ln \frac{k}{aH}  \right) \, .
\vspace{-0.25cm}
$$
For numerical stability reasons we quote constraints on 
\vspace{-0.25cm}
$$
\bar{X}=\frac{\int^{0}_{\ln a_1}{X(a)\,\mathrm{d} \ln a}}{\int^{0}_{\ln a_1}{\mathrm{d} \ln a}} 
\vspace{-0.25cm}
$$ 
We also take the conservative choice of not modifying the gravitational coupling to photons and neutrinos (this currently being unspecified in the PPNC approach). 

Implementation details of these equations in a modified version of the Boltzmann code CLASS \cite{class} is given elsewhere \cite{Thomas_2024}.
Here we put precision constraints on the parameters that appear in them, and their time evolution, using observations covering an unprecedented range of temporal and spatial scales and phenomena.

\vspace{0.25cm}
\textit{Methodology}:
We use sets of Cosmological Data (CD) and Solar System Data (SSD). The CD consist of CMB data from the Planck 2018 data release\footnote{Comprising the low-$\ell$ likelihood, the full TT, EE and TE high-$\ell$ likelihoods with the complete ``not-lite'' set of nuisance parameters, and the Planck lensing potential likelihood (see e.g. \cite{planck_likelihood} and \href{https://wiki.cosmos.esa.int/planck-legacy-archive/index.php/Main_Page}{https://wiki.cosmos.esa.int/planck-legacy-archive/index.php/Main\_Page} for details).} \cite{Planck_2020}, and BAO data\footnote{As implemented in MontePython and used in \cite{2205.05636}, comprising the 6dFGS
($z_{\rm eff}=0.106$) \cite{1106.3366}, the SDSS DR7 MGS
($z_{\rm eff}=0.15$) \cite{1409.3242,1409.3238}, the BOSS
DR12 galaxies ($z_{\rm eff}=0.38, 0.51$) \cite{1607.03155}, the eBOSS
DR16 \cite{2007.08991} LRGs
($z_{\rm eff} = 0.7$) \cite{2007.08993,2007.08994}, ELGs
($z_{\rm eff}=0.85$) \cite{2007.09009, 2007.09008}, quasars ($z_{\rm eff}=1.48$) \cite{2007.08998,2007.08999}, and Lyman-$\alpha$ and Lyman-$\alpha$--quasar cross-correlation ($z_{\rm eff}=2.33$) \cite{2007.08995}.} \cite{1106.3366,1409.3242,1409.3238,1607.03155,2007.08991,2007.08993,2007.08994,2007.09009,
2007.09008,2007.08998,2007.08999,2007.08995}.
The SSD consist of data from the Cassini probe \cite{cassini2003} and Mars ephemeris data \cite{konopliv2011mars}, which is included in the form of Gaussian priors\footnote{Constraints from Cassini \cite{cassini2003} and Mars ephemeris data \cite{konopliv2011mars} give $\gamma=1.000021^{+0.000023}_{-0.000023}$ and $\dot{\alpha}=0.1^{+1.6}_{-1.6}\times10^{-13}\,{\rm yr}^{-1}$ at $a=1$.} on $\dot{\alpha}$ and $\gamma$ at $a=1$. Our constraints are obtained using an MCMC approach via the publicly available MontePython code \cite{audren2013conservative,audren2013monte}, which calls our modified PPNC CLASS code through a Python wrapper. For all runs we use a Metropolis Hastings algorithm and a jumping factor of $2.1$. We set $N_\text{eff}=3.046$, with two massless neutrinos and one with $m=0.06\,$eV. In PPNC the usual nucleosynthesis expression is currently absent, so we vary the seven standard cosmological parameters\footnote{These are the dimensionless density of dark matter $\omega_c$, the dimensionless density of baryons $\omega_b$, the Hubble parameter $H_0$ (in units of km/s/Mpc), the optical depth $\tau$, the amplitude of scalar perturbations $A_s$, the spectral index of scalar perturbations $n_s$, and the primordial helium fraction $Y_{\rm P}$.} $\{ \omega_c, \omega_b, H_0, \tau, \ln(10^{10}A_s), n_s, Y_{\rm P} \}$, together with between $2$ and $4$ PPNC parameters\footnote{The parameter $\gamma({a=1})$ is set to 1 when no Solar System data are used. We note that $\alpha({a=1}) \equiv 1$ by definition. The parameter $n$ will sometimes be fixed to specific values to better understand the physics.} $\{\bar{\gamma}, \bar{\alpha},\gamma({a=1}), n\}$. 

We include Gaussian priors on the nuisance parameters, flat priors on the standard cosmological parameters, and the non-negative priors $0\le\left\lbrace\alpha(a_1),\gamma(a_1)\right\rbrace\le50$. When varying the power-law $n$ we include a flat prior excluding $n>0.25$, as power-laws greater than this only introduce modifications to gravity for a short time in the early universe \cite{Thomas_2024}. 
For the fixed power-law runs we obtain a suitable covariance matrix from exploratory runs, and then run ten chains for each case, continuing until the Gelman-Rubin `$1-R$' convergence test is smaller than $0.01$ for all parameters. Following \cite{Thomas_2024} for the varying power-law cases, we run exploratory chains to find an appropriate covariance matrix that gives a sensible acceptance rate whilst exploring the necessary range of power-law values. We then run $120$ chains (of similar length to those run for the fixed power-law cases) that are individually inspected for converged behaviour; the large number of chains ensures that the parameter space is well explored.

\begin{figure*}
    \centering
\hspace{-1cm}
    \includegraphics[width=0.52\linewidth]{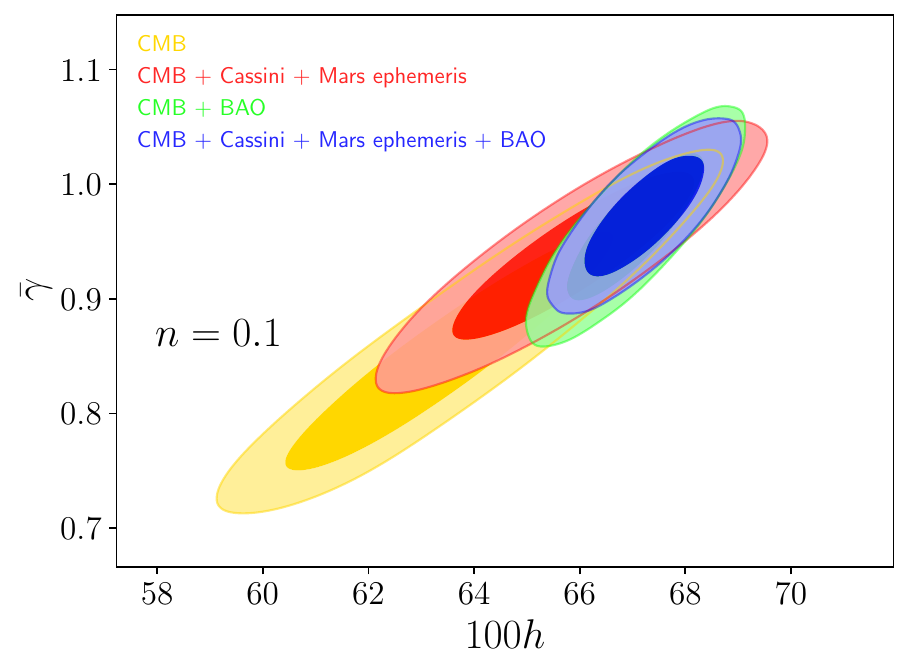}
\hspace{-0.35cm}
    \includegraphics[width=0.52\linewidth]{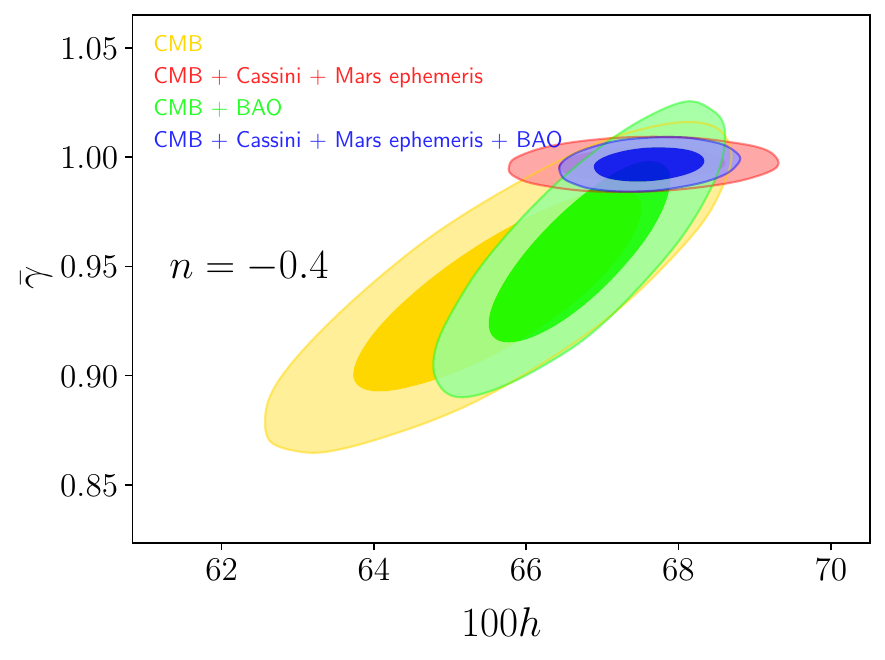}
    \vspace{-0.25cm}
    \caption{2D posteriors on $h = H_0/100\,{\rm km}\,{\rm s}^{-1}\,{\rm Mpc}^{-1}$ and $\bar{\gamma}$ for $n=0.1$ (left) and $n=-0.4$ (right). Results are shown for the Planck data (as in \cite{Thomas_2024}), as well as with the inclusion of SSD and BAO data.}
    \label{fig_powerlawcontours}
\end{figure*}

\vspace{0.25cm}
\textit{Results}:
For fixed power law index $n$, both BAO and SSD improve the CMB-only constraints on the PPNC parameters $\bar{\alpha}$ and $\bar{\gamma}$, as well as the cosmological parameters $H_0$ and $\omega_c$ with which they are degenerate, and decrease departures from $\bar{\alpha}=\bar{\gamma}=1$. These constraints are shown for $n=0.1$ in Table \ref{table_power-lawconstraints} for different combinations of data sets, from which it can be seen that BAO improve constraints more effectively than SSD. This is also shown in the left panel of Figure \ref{fig_powerlawcontours}, where lower values of $\bar{\gamma}$ are curtailed by SSD, and reduced even further by BAO data. For the SSD this is due to the Mars ephemeris constraint, which for fixed $n$ amounts to a constraint on ${\bar{\alpha}}$, and for BAO data is due to stronger constraints on $H_0$. Constraints from SSD become tighter for increasingly negative $n$, whereas BAO data have less effect in this case, as shown in the right-hand panel. 

\begin{table}[b]
\caption{\label{table_power-lawconstraints}
The mean and $95\%$ confidence intervals for $n=0.1$, using 4 combinations of data sets.}
\begin{ruledtabular}
\begin{tabular}{ccccc}
$n=0.1$ & CMB only & CMB+BAO & CMB+SS & CMB+BAO+SS \\ 
\hline
$100h$ & $63.7^{+4.0}_{-3.9}$ & $67.1^{+1.7}_{-1.7}$ & $65.6^{+3.1}_{-3.1}$ & $67.2^{+1.5}_{-1.5}$ \\
$\bar{\gamma}$ & $0.86_{-0.13}^{+0.13}$ & $0.96^{+0.09}_{-0.09}$ & $0.94^{+0.10}_{-0.10}$ & $0.97^{+0.07}_{-0.07}$ \\ 
\end{tabular}
\end{ruledtabular}
\end{table}

\begin{figure*}
 \vspace{-0.25cm}
    \includegraphics[width=0.9\linewidth]{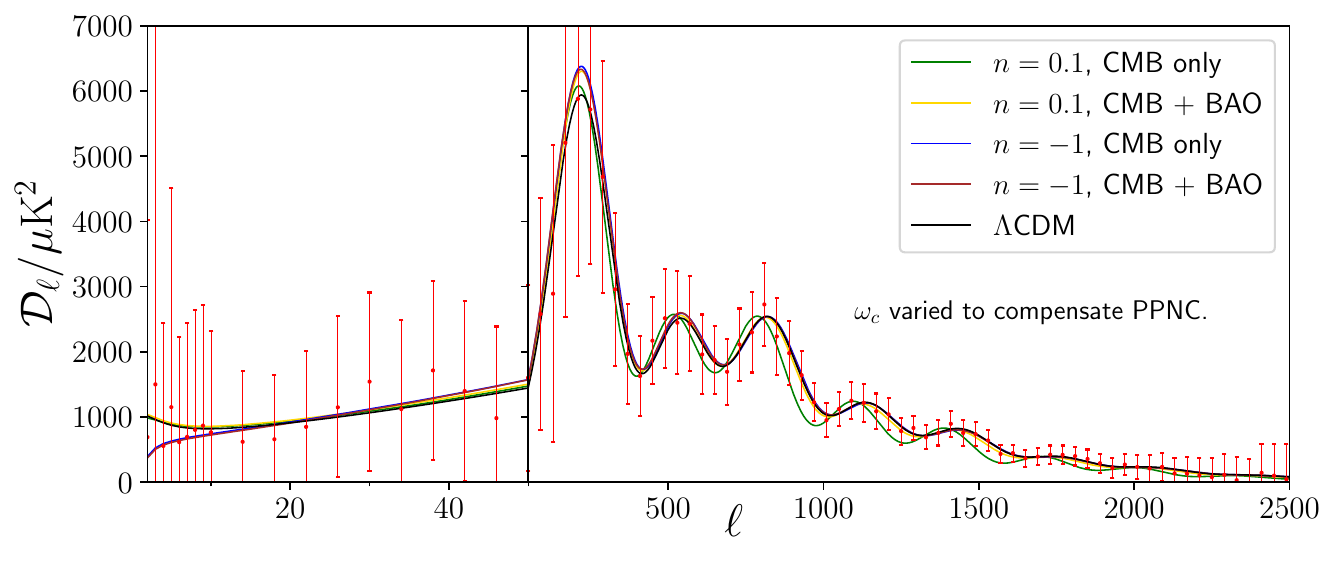}
    \vskip -1ex
    \vspace{-0.7cm}
    \hspace{0.05cm}
    \includegraphics[width=0.9\linewidth]{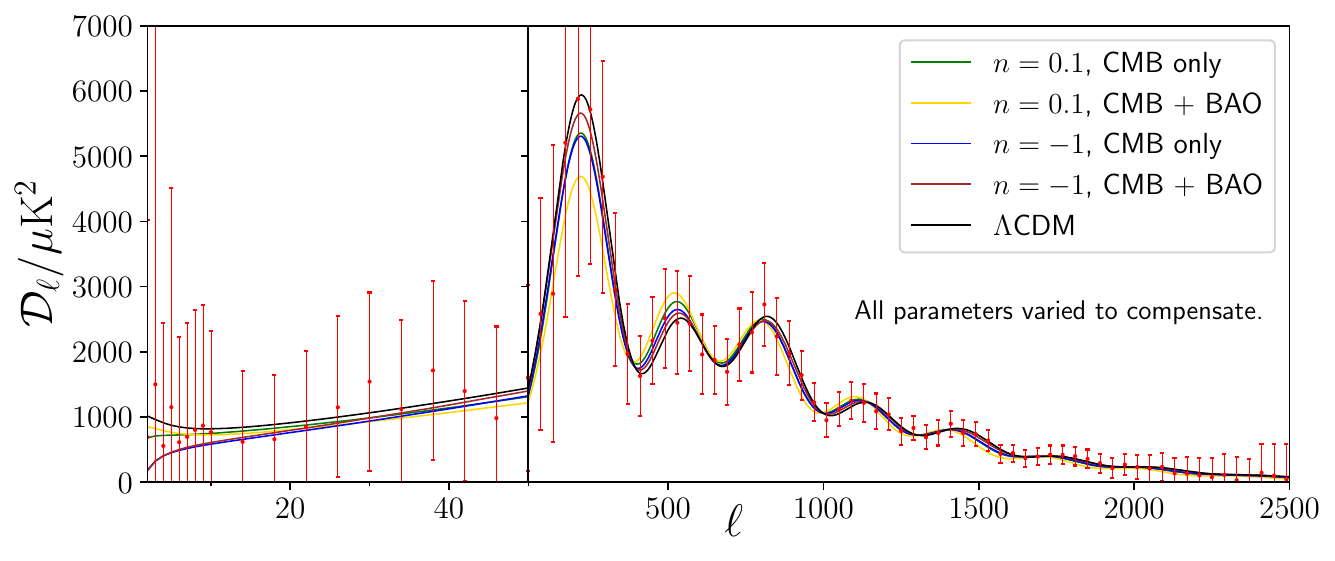}
    \vspace{-0.25cm}
    \caption{CMB power spectra for central parameter values. {\it Top panel}: all  cosmological parameters except $\omega_c$ take their central $\Lambda$CDM values, and $\omega_c$ takes its central value from the PPNC chains. {\it Lower panel}: all cosmological parameters are adjusted. For $n=0.1$ and $n=-1$ we have $\bar{\gamma}=0.86 (0.96)$ and $0.91 (0.93)$ for CMB(CMB+BAO), respectively. The Planck error bars, and the deviations of all curves from $\Lambda$CDM, have been multiplied by $5$.}
    \label{fig_cmbcls}
\end{figure*}

The effects of including the BAO data are explored further in Figure \ref{fig_cmbcls}. For CMB alone, the case with $n=0.1$ has large deviations from GR that can be compensated by using $H_0$ to adjust the CMB peak locations, whereas for $n=-1$ the smaller deviations from GR are most easily compensated by varying $\omega_c$ (as expected from the role $\gamma$ plays in the background and perturbation equations). The BAO data prevents this compensation for larger values of $n$, making them a worse fit to the data (even for much smaller deviations from GR). This is demonstrated by the $n=0.1$ curve with BAO data having the worst fit to the data of the compensated curves, despite having the smallest deviation from $\bar{\alpha}=\bar{\gamma}=1$ (c.f. the $n=0.1$ curve with CMB data only, which fits well despite being much further from GR). For CMB data only, the likelihood values for $n=0.1$ and $n=-1$ are similar, whereas the likelihood is better for increasingly negative $n$ when including BAO data. This latter behaviour is due to better compensation,
which happens because the PPNC parameters stay closer to constant (which is simpler behaviour to accommodate and also removes any effects from the time derivatives of $\gamma(a)$ at early times). The $n=-1$ curves, particularly the curve that includes BAO data, show signs of performing better than the central $\Lambda$CDM values: notably around the rise of the second and fourth peaks, the descent of the third peak, and around data points 8-11 at low $\ell$. It is not, however, significant enough for $\Lambda$CDM to be disfavoured at any appreciable confidence level. 

\begin{table}[b]
\caption{\label{table_powerlawconstraints}
The mean and $95\%$ confidence intervals for 3 combinations of data sets, when $n$ is not fixed. 
}
\begin{ruledtabular}
\begin{tabular}{cccc}
 & CMB only & CMB+SS & CMB+BAO+SS \\ 
\hline
$\bar{\gamma}$ & $0.90^{+0.14}_{-0.15}$ & $0.96^{+0.13}_{-0.13}$ & $0.97^{+0.10}_{-0.11}$  \\ 
$\bar{\alpha}$ & $0.89^{+0.17}_{-0.18}$ & $0.96^{+0.16}_{-0.16}$ & $0.97^{+0.13}_{-0.14}$ \\ 
$10\omega_c$ & $1.21^{+0.04}_{-0.04}$  & $1.19^{+0.03}_{-0.03}$ & $1.19^{+0.02}_{-0.02}$ \\
$100h$ & $65.1^{+3.9}_{-4.1}$  & $66.8^{+2.6}_{-2.9}$ & $67.4^{+1.4}_{-1.4}$ \\
$n$ & $0.084^{+0.17}_{-0.34}$  & $0.13^{+0.12}_{-0.29}$ & $0.12^{+0.13}_{-0.41}$ \\
\end{tabular}
\end{ruledtabular}
\end{table}



\begin{figure*}
    \centering
    \includegraphics[width=0.75\linewidth]{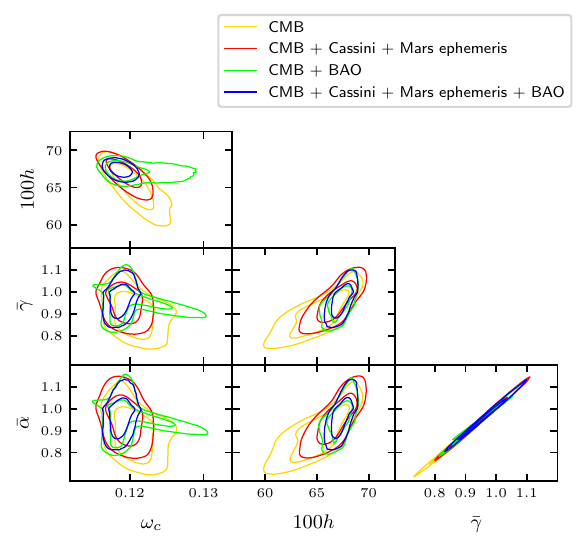}
    \vspace{-0.25cm}
    \caption{2D $0.15\,\sigma$-smoothed posteriors for $\omega_c$, $H_0$, $\bar{\gamma}$ and $\bar{\alpha}$, for the case with varying $n$.}
    \vspace{-0.15cm}
    \label{fig_varyingpowercontours0}
\end{figure*}

\begin{figure}
 \hspace{-0.5cm}
    \includegraphics[width=1.05\linewidth]{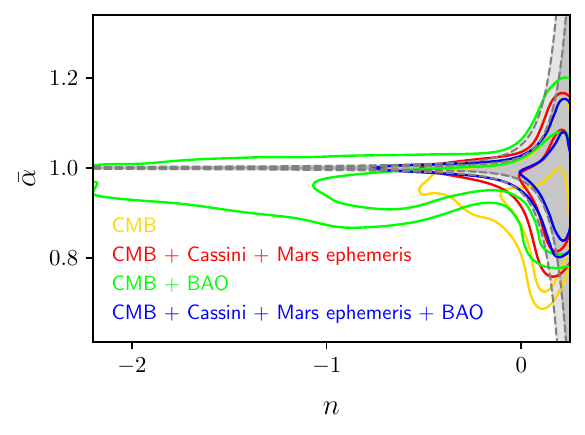}
      \vspace{-0.25cm}
    \caption{$0.15\,\sigma$-smoothed 2D posteriors in the $\bar{\alpha}-n$ plane, showing in grey the $68\%$ and $95\%$ contours corresponding to the Solar System prior on $\dot{\alpha}(a=1)$ coming from the data on the ephemeris of Mars.}
    \label{fig_varyingpowercontours}
\end{figure}

For more general cases, in which $n$ is not fixed to a pre-specified value, the key constraints are given in Table \ref{table_powerlawconstraints} (we do not show $\Lambda$CDM parameters as they change only minimally compared to the CMB-only case). The inclusion of SSD tightens the constraints, with the allowed ranges of $H_0$, $\bar{\alpha}$ and $\bar{\gamma}$ being reduced and shifted back towards the $\Lambda$CDM values for larger values of $n$, as shown by the difference between the yellow and red contours in Figure \ref{fig_varyingpowercontours0}, and as expected from the cases with $n$ fixed. The shape of the Mars ephemeris constraint in the $\bar{\alpha}$-$n$ plane is shown in grey in Figure  \ref{fig_varyingpowercontours}, pushing the CMB-only constraints back towards the GR values of $\bar{\alpha}=\bar{\gamma}=1$, along with resultant effects on $H_0$ and $\omega_c$ (due to their degeneracies with the $\bar{\alpha}$ and $\bar{\gamma}$). These contours come from converting the constraint on $\dot{\alpha}(a=1)$ to constraints on $\bar{\alpha}$ for a given $n$, while holding $H_0$ at its $\Lambda$CDM value\footnote{The variance in $\bar{\alpha}$ is overwhelmingly dominated by the variance in $\dot{\alpha}(a=1)$, so the effect of varying $H_0$ is negligible here.}. Considering CMB+BAO only more strongly limits the allowed ranges of $H_0$, $\bar{\alpha}$ and $\bar{\gamma}$, while also allowing for a much wider range of values of $n$ (due to the reduced ability to compensate larger values of $n$, and the small improvement in likelihoods for increasingly negative $n$, as discussed above)\footnote{We do not include CMB+BAO in the varying $n$ case in Table \ref{table_powerlawconstraints} as we were not able to demonstrate that the chains were converged to a high degree of robustness for $\omega_c$ and $n$. However, we include the contours in the plots as the qualitative behaviour is clear and provides additional elucidation of the physical effects and degeneracies, thus clarifying the CMB+SS+BAO results.}. This large tail for $n$ is a 3-way degeneracy between $n$, $\omega_c$ and $\bar{\gamma}$ due to the compensation of the effects  of introducing the PPNC parameters. As a result, the constraints on $\omega_c$ and $n$ worsen, whereas the others improve. These effects can be seen in Figure \ref{fig_varyingpowercontours0}. The wider range of allowed $n$ values, and the small bias towards PPNC parameter values below one for more negative values of $n$, can be seen in Figure \ref{fig_varyingpowercontours}.

 Including both SSD and BAO data yields the tightest constraints on all parameters except $n$.\footnote{The BAO data and SSD act to push the preferred range of $n$ in opposite directions, and adding BAO data only extends the posterior by a small amount compared to CMB+SSD due to the strength of the Mars ephemeris constraint.} The average deviation from GR over cosmic time is constrained to be less than $10\%$, with $\bar{\gamma}=0.97^{+0.05}_{-0.05}$ and $\bar{\alpha}=0.97^{+0.06}_{-0.07}$ at the $68\%$ confidence level. Generally, the parameter degeneracies follow the shapes of the CMB+SSD case, with the inclusion of BAO data primarily curtailing the region around $n=0.1$ where smaller values of $\bar{\alpha}$ and $\bar{\gamma}$ are compensated by smaller values of $H_0$.

Figure \ref{fig_varyingpowercontours0} shows there is a strong degeneracy between the PPNC parameters in all cases, with the difference between $\bar{\alpha}$ and $\bar{\gamma}$ over cosmic time constrained to be less than $3\%$ ($0.0004^{+0.0096(0.0265)}_{-0.0124(0.0251)}$ at $68\%$ ($95\%$) confidence for CMB+SSD+BAO). This means that observers performing local tests of gravity at different cosmic times would typically find local gravitational physics to be compatible with GR, albeit with different values for $G$, even though there is still substantial variation allowed in the time evolution of $\alpha$ and $\gamma$. When all data sets are combined, both $\dot{\alpha}$ and $\dot{\gamma}$ at $a = 1$ are constrained to be $0.4^{+0.6}_{-0.8}\times 10^{-13}$ yr$^{-1}$ at $68\%$ confidence, which is an improvement of more than a factor of two compared to the Mars ephemeris constraint on $\dot{\alpha}$ alone, and provides a first constraint on $\dot{\gamma}$.

\vspace{0.25cm}
\textit{Discussion}:
We have demonstrated the power of a new framework for precision tests of gravity; the Parametrized Post-Newtonian Cosmology (PPNC). This is the cosmological extension of the classic PPN approach, and allows one to combine Solar System and cosmological data to constrain a single set of gravitational parameters. Such combinations significantly improve constraints on GR over cosmological time, whilst clearly showing the freedoms that remain. This is the first study that directly combines Solar System and cosmological data to constrain a single coherent set of theory-agnostic gravity parameters, opening the door for precision tests of gravity by combining data from all length and time scales.

The average value of the PPNC parameters over the age of the universe is consistent with GR, with deviations constrained to be within 10\%: $\bar{\gamma}=0.97^{+0.05}_{-0.05}$ and $\bar{\alpha}=0.97^{+0.06}_{-0.07}$. Additionally, we have shown that the difference between $\bar{\alpha}$ and $\bar{\gamma}$ is constrained to be around the percent level, meaning that observers performing local tests of gravity at any cosmic time would infer GR to be the correct theory of gravity (just with a different value of Newton's constant, $G$). The PPNC framework allows us to constrain parameters that are not constrained by Solar System observables alone, such as $\dot{\gamma}(a=1)=0.4^{+0.6}_{-0.8}\times10^{-13}$ yr$^{-1}$, while tightening Solar System constraint on $\dot{\alpha}(a=1)$.

The cosmological datasets that can be used with this framework could be expanded by applying it within the Effective Field Theory of Large Scale Structure \cite{eftoflss} (to include, e.g., the DESI data \cite{desi}), or by modifying N-body simulations to include the modified equations for non-linear structure formation (the resulting predictions of which would allow the use of much larger amounts of data from Euclid and/or LSST). It would also be interesting to extend this work to more general forms for $\gamma(a)$ and $\alpha(a)$ using techniques such as Gaussian Processes or Genetic Algorithms. Our constraints do not change substantially depending on whether BAO, SSD or both are used, particularly for $H_0$ and the PPNC parameters, but this may well be a result of our choice of power-law behaviour. There is, however, a degeneracy between $\omega_c$ and $n$ for CMB+BAO that is removed by adding Solar System data. As such, the constraint on $\omega_c$ is expected to be dependent on the choice of a power-law time dependence. More generally, since the Solar System constraining power is concentrated at one instant in time, which is different to the times that are relevant for cosmological observables, we expect that more general forms of the time dependence would further increase the importance of combining the different datasets.

The PPNC framework demonstrated in this work marks a paradigm shift in precision tests of GR. It will enable us to make optimal use of ongoing and upcoming experiments for constraining gravity, by combining data from a wide range of times, scales and phenomena into constraints on a single unified set of parameters.

\vspace{0.25cm}
\textit{Acknowledgements} This research used QMUL's Apocrita HPC facility, supported by QMUL ITS Research (http://doi.org/10.5281/zenodo.438045). We acknowledge the assistance of the ITS Research Team at Queen Mary University of London. DBT thanks Phil Bull for useful discussions. DBT, TA and TC acknowledge support from the Science and Technology Facilities Council (STFC, grant numbers ST/P000592/1 and ST/X006344/1). We acknowledge the use of GetDist \cite{getdist}.

\bibliography{apssamp}

\end{document}